\documentclass[manuscript,screen,authorversion,nonacm]{acmart}
\pdfoutput=1

\usepackage{url}

\AtBeginDocument{%
  \providecommand\BibTeX{{%
    \normalfont B\kern-0.5em{\scshape i\kern-0.25em b}\kern-0.8em\TeX}}}

\setcopyright{none}
\copyrightyear{2022}
\acmDOI{}
\acmConference[]{}{}{}
\acmBooktitle{}
\acmPrice{}
\acmISBN{}

\begin{document}

\title{AI Ethics Statements}
\subtitle{Analysis and lessons learnt from NeurIPS Broader Impact Statements}

\author{Carolyn Ashurst}
\authornote{These authors contributed equally to this research.}
\email{cashurst@turing.ac.uk}
\affiliation{%
  \institution{Alan Turing Institute}
}

\author{Emmie Hine}
\authornotemark[1]
\email{emmie.e.hine@gmail.com}
\affiliation{%
  \institution{Oxford Internet Institute}
}

\author{Paul Sedille}
\authornotemark[1]
\email{paul.sedille@gmail.com}
\affiliation{%
  \institution{Harvard Kennedy School}
}

\author{Alexis Carlier}
\email{alexis.carlier@philosophy.ox.ac.uk}
\affiliation{%
  \institution{Future of Humanity Institute}
}

\renewcommand{\shortauthors}{Ashurst, Hine and Sedille, et al.}

\begin{abstract}
Ethics statements have been proposed as a mechanism to increase transparency and promote reflection on the societal impacts of published research. In 2020, the machine learning (ML) conference NeurIPS broke new ground by requiring that all papers include a broader impact statement. This requirement was removed in 2021, in favour of a checklist approach. The 2020 statements therefore provide a unique opportunity to learn from the broader impact experiment: to investigate the benefits and challenges of this and similar governance mechanisms, as well as providing an insight into how ML researchers think about the societal impacts of their own work. Such learning is needed as NeurIPS and other venues continue to question and adapt their policies. 
To enable this, we have created a dataset containing the impact statements from all NeurIPS 2020 papers, along with additional information such as affiliation type, location and subject area, and a simple visualisation tool for exploration.
We also provide an initial quantitative analysis of the dataset, covering representation, engagement, common themes, and willingness to discuss potential harms alongside benefits. We investigate how these vary by geography, affiliation type and subject area.
Drawing on these findings, we discuss the potential benefits and negative outcomes of ethics statement requirements, and their possible causes and associated challenges. These lead us to several lessons to be learnt from the 2020 requirement: (i) the importance of creating the right incentives, (ii) the need for clear expectations and guidance, and (iii) the importance of transparency and constructive deliberation.
We encourage other researchers to use our dataset to provide additional analysis, to further our understanding of how researchers responded to this requirement, and to investigate the benefits and challenges of this and related mechanisms.
\end{abstract}

\maketitle

\section{Introduction}

In response to the increasing recognition of harms resulting both from deployed systems \citep{noble2018algorithms, benjamin2019race, eubanks2018automating, o2016weapons, barocas2016big, obermeyer2019dissecting, zuboff2019age} and research outputs \citep{FATE_tutorial, abolish_techtoprison, research_shoudnt, microsoft_disinformation}, some have called on the machine learning (ML) research community, including ML conferences and journals, to do more to promote ethical research \citep{hecht_its_2018}. This has led to a range of initiatives such as codes of ethics \citep{gotterbarn2017acm}, ethics committees \citep{emnlp_call_2020} and ethics review boards \citep{bernstein2021esr}.

Perhaps the most significant change relating to the publication of papers was introduced by the NeurIPS conference. The NeurIPS program chairs announced that all authors submitting to the 2020 conference must include a broader impact section, in which authors should discuss “the broader impact of their work, including possible societal consequences -- both positive and negative” \citep{NeurIPS_getting_2020, neurips_call_2020}. In addition, a new ethics review process was incorporated into the peer review process. Technical reviewers could flag papers for potential ethical concerns, to be reviewed by a team of ethics experts \citep{lin_what_2020}. These initiatives provoked both praise and criticism, showing a lack of consensus on whether these mechanisms should be adopted \citep{hecht_its_2018, johnson_neurips_2020, pai_pub_norms_2021}, and how they should be operationalised \citep{prunkl2021institutionalizing}.

A year later, in their first blog post for the 2021 conference, the 2021 chairs described how they had reviewed the broader impact requirement, taking into account an author survey, views expressed at the Broader Impacts Workshop \citep{nbiair_git_2020}, similar efforts in other communities, and discussions with the ML community and beyond \citep{intro_checklist}. The chairs stated that ``authors want both more guidance around how to perform machine learning research responsibly and more flexibility in how they discuss this in their papers''. In light of this, the chairs removed the requirement to include a separate broader impact section, and introduced a new checklist for authors to include in their submitted paper \citep{neurips_call_2021}. The checklist questions relate to a range of responsible research practices, including reproducibility and scientific best practice. Regarding societal impacts, the authors are asked “Did you discuss any potential negative societal impacts of your work?”, and are offered some high level guidance on what to consider. While it is no longer mandatory to include a separate impact statement, the guidance suggests that a discussion of negative societal impacts is expected of researchers, though how to incorporate this into their paper is left to the authors \citep{neurips_checklist_2021}.

The 2020 broader impact statements thus represent a unique opportunity to investigate the benefits and challenges associated with ethics statement requirements.
Such work can help inform decisions around ethics statements for other contexts including funding applications, organisational approval, or for other publication venues who may be considering introducing such requirements, or who have already done so (such as EMNLP \cite{emnlp_call_2021}). Lessons may also be applied to the NeurIPS checklist approach (since a discussion of negative impacts remains a component), and future NeurIPS policy. 

\subsection{Summary of findings}
\label{sec:findings}
Our ultimate goal is to better understand the implications of impact statement requirements and related mechanisms, including their benefits, risks and challenges. As a step towards this goal, our main contributions are: (i) an open source dataset of impact statements from all NeurIPS 2020 papers, along with additional information such as affiliation type, location and subject area \citep{BIS_github}, (ii) a simple visualisation tool for exploration of the dataset \citep{flourish_visualisation}, (iii) an initial quantitative analysis of the dataset, in which we explore representation, engagement, themes and valence (\S~ \ref{sec:analysis}) and (iv) a discussion of benefits, risks and challenges evidenced by these findings, from which we draw several lessons learnt (\S~ \ref{sec:discussion}).

Based on our analysis our main findings are as follows.

\paragraph{Voices represented} There is concentration of authors of papers, and therefore impact statements, along both geographic (\S~\ref{sec:loc}) and institutional (\S~\ref{sec:top_aff}) lines. Authors are concentrated in North America (67\% of papers have at least one North American affiliation), followed by Europe (29\%) and Asia (27\%), and a significant number of affiliations are concentrated in a handful of institutions, such as Google (13\% have at least one Google affiliation), Stanford (7\%) and Microsoft (5\%). We also note the large overlap between industry and academic affiliations, with around a third of papers having affiliations from both industry and academia (\S~\ref{sec:aff_type}).

\paragraph{Engagement with broader impacts} We find high variation in engagement as measured by statement length (\S~\ref{sec:length}) and opt-out rates (\S~\ref{sec:opt-outs}). While the average statement length is only 169 words (around 7 sentences), the distribution of lengths has a long tail -- the longest statement containing over 4000 words. Around 10\% of papers choose to effectively opt out of writing a statement, for example by stating that it is ``not applicable''. These vary greatly by subject area; \textit{Theory} and \textit{Optimization} had the highest opt-out rates (25\% and 24\%), and \textit{Applications} and \textit{Social Aspects of ML} had the lowest (2\% and 1\%).

\paragraph{Themes} We find evidence that certain well established topics were common to many impact statements, including privacy, fairness, robustness and safety (\S~\ref{sec:themes}). The most frequent words associated with application settings were \textit{medical}, \textit{robots} and \textit{science}.

\paragraph{Valence} We find evidence that authors tend to discuss more positive aspects compared to negative aspects in their statements (\S~\ref{sec:val}). On average, statements included synonyms of \textit{positive} and \textit{strength} 4.6 and 1.3 times respectively, with synonyms of \textit{negative} and \textit{limitation} occurring 3.6 and 0.6 times.\\

These findings highlight some of the challenges and issues associated with broader impact requirements, such as incentives to downplay negative impacts, lack of incentives to engage deeply with the task, and a concentration of perspectives along geographic and institutional lines. Despite these challenges, our findings also highlight several benefits of such requirements. Some authors did engage with the task thoughtfully, with some taking the opportunity to thoroughly investigate potential ethical issues, indicating that the requirement can promote reflection and awareness raising. The statements also give us a sense of which issues are widely recognised by authors (such as privacy and fairness), which can help us understand which issues are comparatively neglected. Given this was the first year of the requirement, with lightweight guidance and few examples for authors to refer to, we cannot judge its full potential. Even so, there are several lessons we can draw from this initial attempt. These include (i) the importance of creating the right incentives, (ii) the need for clear expectations and guidance, and (iii) the importance of transparency and constructive deliberation. As the new checklist requirement for NeurIPS 2021 papers asks authors whether they have considered potential negative societal impacts, the community should continue to monitor how researchers respond, and to reflect on the utility of such requirements.

\subsection{Related work}
Before the final statements were available, work to understand how researchers had responded to the NeurIPS impact statement requirement included an analysis of preprints available before the conference \citep{boyarskaya2020overcoming}, a survey of researcher attitudes towards the requirement \citep{abuhamad2020like}, and reflections on potential benefits and challenges based on related mechanisms \citep{prunkl2021institutionalizing}. To date the most significant investigation into the final statements is provided by \citet{nanayakkara2021unpacking}, who provide a qualitative thematic analysis of a sample of 300 statements. They identify several themes related to how consequences are expressed (such as valence, specificity and uncertainty), the areas of impacts expressed (such as privacy, labor, the environment, efficiency and robustness), and researcher’s recommendations for mitigations. Our work complements theirs by providing a dataset and analysis of all 1898 statements with additional information (such as affiliation location and type), code and visualisation tools for further analysis, and an initial analysis that both builds on some of their identified themes (such as valence and areas of impacts), and asks complementary questions (such as those around engagement, and investigating how engagement, themes and valence differ by subject area, geography and affiliation type). We also discuss evidence of benefits, negative outcomes, causes and challenges from these findings, in order to summarise recommendations for future self-governance mechanisms. 
\section{Data}
We obtained the manuscript pdfs for accepted papers from the NeurIPS 2020 proceedings website \citep{neurips_proceedings_2020}, which contains 1898 papers.\footnote{In November 2020 the website contained 1899 papers. One paper was later taken down from the site and has therefore been removed from our dataset, which now includes 1898 papers.} We converted the pdfs to XML and extracted 
the title and impact statement section \citep{pdf-scraper}. We appended this dataset with information about paper subject area, author names,  affiliations, affiliation type and affiliation institution locations, as follows. Primary and secondary subject area, as selected by authors on submission, were supplied to us by the NeurIPS programme chairs \citep{data_subject_areas_google_sheet}. Author names \citep{neurips_proceedings_2020} and affiliations \citep{neurips_2020_affiliations} were obtained from separate scrapes of the NeurIPS papers. Each affiliation was tagged with a location and type (industry or academia) based on \citep{neurips_2020_locations} and \citep{affiliation_type_google_sheet} respectively. Further details on dataset generation, and the assumptions and limitations of our dataset can be found in our code documentation \citep{data_generation_github}.
The resulting dataset and our code can be found in the github repository \citep{BIS_github}. We also created a simple visualisation tool, for convenient exploration of the dataset, which can be found on Flourish \citep{flourish_visualisation}.
\section{Analysis}
\label{sec:analysis}
We divide our analysis into four categories: representation, engagement, themes and valence.

\subsection{Representation}
\label{sec:voices}

In this section we explore whose voices are represented in the 2020 broader impact statements, by investigating the affiliation types, organisations, geography and subject area of the paper authors.

\subsubsection{Affiliation type}
\label{sec:aff_type}

We classified each affiliation listed as academia or industry, based on \citep{affiliation_type_google_sheet}. Papers were then categorised as \textit{academia} (if all affiliations were from academia), \textit{industry} (if all affiliations were from industry), or \textit{mixed} (if the paper had both academic and industry affiliations listed).
According to this categorisation, 1163 (61\%) of papers were authored solely by academic authors, 122 (6\%) of papers were authored solely by authors with industry affiliations, and 613 (32\%) had authors from both academia and industry -- see Table ~\ref{tab-type}.

\begin{table}[htbp!]
  \caption{Academia versus industry}
  \label{tab-type}
  \centering
  \begin{tabular}{lll}
  \toprule
    Affiliation type     & Number of papers     & Percentage \\
    \midrule
    Academia & 1163 & 61\%     \\
    Industry & 122  & 6\%      \\
    Mixed     & 613  & 32\%     \\
    \bottomrule
  \end{tabular}
\end{table}

\subsubsection{Most represented affiliations}
\label{sec:top_aff}
In Table ~\ref{tab-affil} we show all institutions associated with at least 30 accepted papers. The most common affiliations were 
Google (240 papers, 13\%), Stanford (141, 7\%), Microsoft (98, 5\%), MIT (92, 5\%) and UC Berkeley (79, 4\%).
Together, over 28\% of papers have at least one author affiliated with these five institutions. 
Over 16\% of papers have an Alphabet affiliation (such as Google or DeepMind). 

\begin{table}
  \caption{Most represented affiliations - \url{https://public.flourish.studio/visualisation/6152118}}
  \label{tab-affil}
	\begin{minipage}{0.5\linewidth}
  \centering
  \begin{tabular}{lll}
  \toprule
    Institution    & Number of papers     & Percentage \\
    \midrule
Google$^\ast$  &	240	& 13\% \\
Stanford University	& 141	& 7\% \\
Microsoft &	98 &	5\% \\
MIT	& 92	& 5\% \\
UC Berkeley	&79 &	4\% \\
DeepMind &	76	&4\% \\
Tsinghua University	&74&	4\%\\
Princeton University &	63 &	3\% \\
Columbia University	&52&	3\% \\
ETH Zürich	& 51	& 3\% \\
Harvard University &43&	2\% \\
Cornell University	&41&	2\% \\
Facebook &	38 &	2\% \\
Peking University	&38&	2\%\\
UCLA	&38&	2\% \\
IBM	& 38&	2\% \\
EPFL &	30&	2\% \\
UIUC &	30&	2\% \\
    \bottomrule
    ${}^\ast$\textit{including ``Google DeepMind''}
  \end{tabular}
	\end{minipage}\hfill
	\begin{minipage}{0.45\linewidth}
		\centering
  \includegraphics[width=\columnwidth]{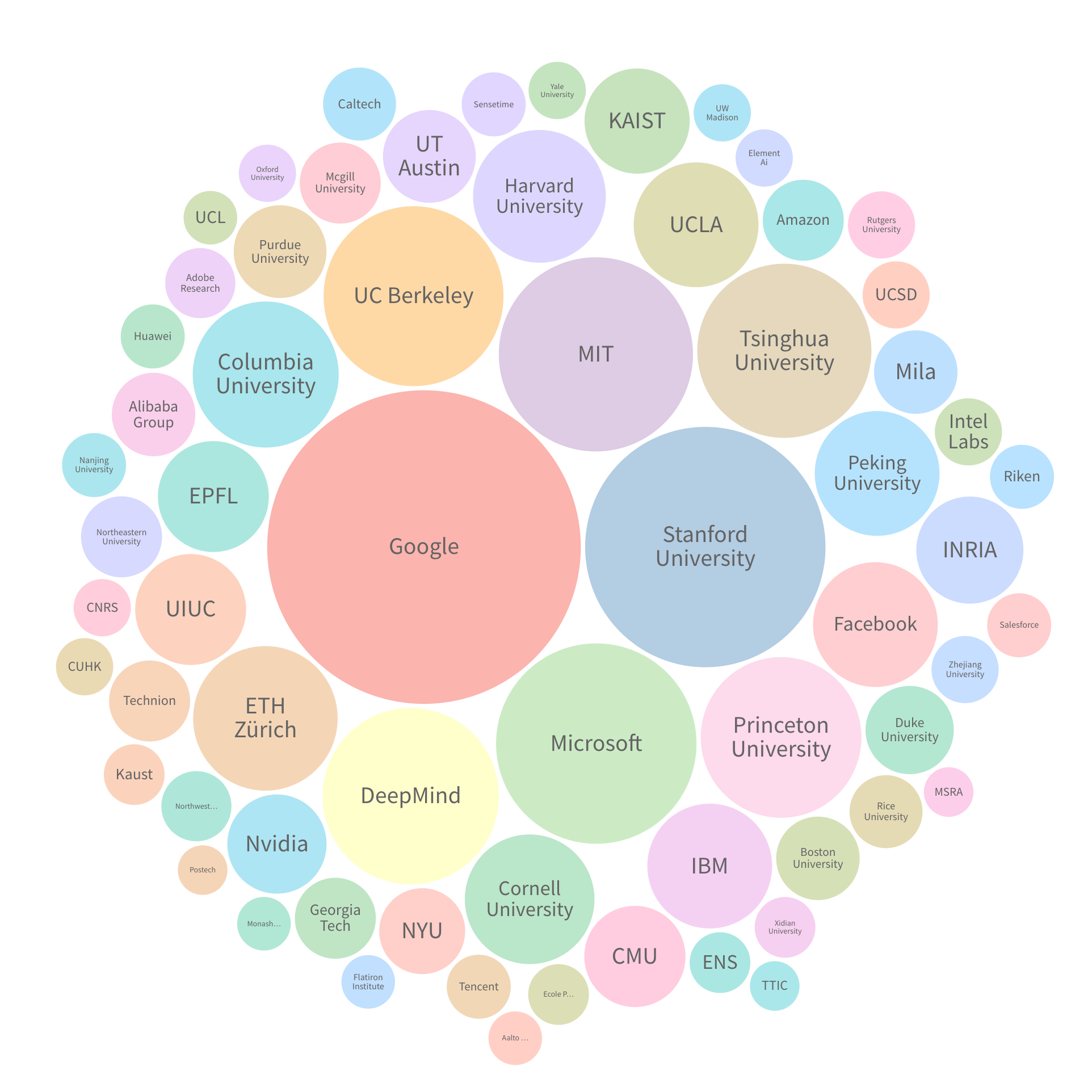}
	\end{minipage}
\end{table}

\subsubsection{Location}
\label{sec:loc}
We grouped affiliation countries into continents -- see Table ~\ref{tab-regions}. We found that around two-thirds of papers had at least one North American affiliation (1279 papers, 67\%), almost a third had a European affiliation (557, 29\%) and likewise for Asian affiliations (551, 27\%). There were only 58 papers with an affiliation from Oceania (3\%), 10 for South America (0.5\%) and only three from Africa (0.2\%). Note that the percentages do not sum to 100\% since papers can have authors from more than one region. US affiliations make up the majority of those from North America -- 65\% of all papers have a US affiliation.

A heatmap showing the number of papers with at least one author affiliation associated with each country is shown alongside Table ~\ref{tab-regions}. The ten countries with the most associated papers were the US (1229 papers, 65\%), China (295, 16\%), the UK (222, 12\%), Canada (110, 6\%), France (106, 6\%), Switzerland (94, 5\%), Germany (88, 5\%), South Korea (59, 3\%), Australia (58, 3\%) and Israel (55, 3\%).

\begin{table}
  \caption{Regions - Number and percentage of papers containing at least one affiliation from each region and number of papers associated with each country - \url{ https://public.flourish.studio/visualisation/6084514}}
  \label{tab-regions}
	\begin{minipage}{0.5\linewidth}
  \centering
  \begin{tabular}{lll}
  \toprule
    Region     & Number of papers     & Percentage \\
    \midrule
    North America & 1279 & 67\%     \\
    Europe & 557 & 29\%     \\
    Asia & 551 & 27\%     \\
    Oceania & 58 & 3\%     \\
    South America & 10 & 0.5\%     \\
    Africa & 3 & 0.2\%     \\
    \bottomrule
  \end{tabular}
	\end{minipage}\hfill
	\begin{minipage}{0.5\linewidth}
		\centering
        \includegraphics[width=\columnwidth]{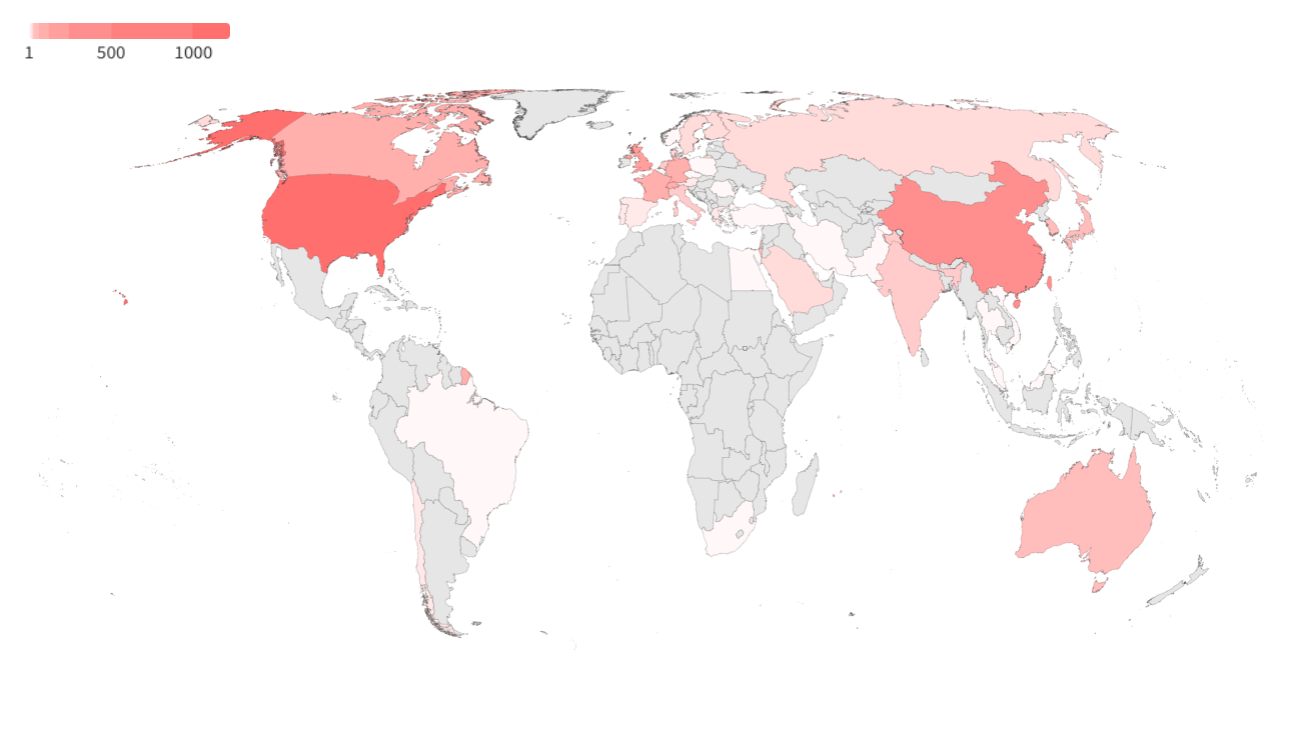}
	\end{minipage}
\end{table}

\subsubsection{Subject area}
\label{sec:subject}
There were 162 granular primary subject areas represented, but these are grouped into ten main categories, such as \textit{Deep Learning} or \textit{Theory}. Table ~\ref{tab-subject} shows the number of papers for each category. We note that while there are a range of subject areas represented, a large proportion of papers have primary subject areas indicating work towards the theoretical end of the spectrum, such as \textit{Theory} (195 papers, 10\%), \textit{Optimization} (125, 7\%), and \textit{Algorithms} (504, 27\%). Those in method based subject areas such as \textit{Deep Learning}, \textit{Reinforcement Learning} and \textit{Probabilistic Methods} likely span the theory-application spectrum, including many that focus on general purpose techniques that may be applied to a range of applications.

\begin{table}[htbp!]
  \caption{Primary subject area}
  \label{tab-subject}
  \centering
  \begin{tabular}{lll}
  \toprule
    Primary subject area & Number of papers & Percentage \\
    \midrule
    Algorithms & 504 & 27\%     \\
    Deep Learning & 325 & 17\%     \\
    Applications & 283 & 15\%     \\
    Theory & 195 & 10\%     \\
    Reinforcement learning and planning & 183 & 10\%     \\
    Optimization & 125 & 7\%     \\
    Probabilistic methods & 125 & 7\%     \\
    Social aspects of machine learning & 77 & 4\%     \\
    Neuroscience and cognitive science & 69 & 4\%     \\
    \multicolumn{1}{p{5cm}}{\raggedright Data, challenges, implementations, \\ and software }       & 12 & 1\%     \\
    \bottomrule
  \end{tabular}
\end{table}

\subsection{Engagement with broader impact statements}
\label{sec:engagement}

\subsubsection{Opt-outs}
\label{sec:opt-outs}
While it was mandatory to include a broader impact section, the NeurIPS template stated that ``If authors believe this is not applicable to them, authors can simply state this'' \citep{neurips_template_2020}. In the FAQs authors were told that ``if your work is very theoretical or is general enough that there is no particular application foreseen, then you are free to write that a Broader Impact discussion is not applicable'' \citep{neurips_faq_neurips_2020}. Here we investigate how many authors chose to “opt out” in this way.

We made the assumption that any statement of more than 60 words constituted an attempt to include a statement, and labelled these as opt-in. We manually labelled all statements of 60 words or less as opt-out or opt-in depending on their content; there were 361 such statements. While many statements were clearly an opt-out, there were many ambiguous examples. For example, some authors imply that an impact statement is not applicable, but go on to mention real world applications (e.g.\ \citep{ging2020coot}), or detailed descriptions of the impact of their work on the field (e.g.\ \citep{meng2020sufficient}). Other ambiguous examples include stating that they do not anticipate any impacts or ethical aspects that are not ``well understood by now'' \citep{moulos2020finite}, or mentioning that their method ``brings risk'', without elaborating \citep{neyshabur2020being}. We labelled 74 statements as ambiguous; 45 of which we decided on balance to include as opt-outs.

In total, we labelled 185 statements (10\%) as opt-outs -- see Table ~\ref{tab-opt}.

\begin{table}[htbp!]
  \caption{Opt-outs}
  \label{tab-opt}
  \centering
  \begin{tabular}{lll}
  \toprule
         & Number of papers     & Percentage \\
    \midrule
    Opt-out & 185 & 10\%     \\
    Opt-in & 1713  & 90\%      \\
    \bottomrule
  \end{tabular}
\end{table}

\begin{figure}[htbp!]
  \centering
  \includegraphics[width=0.6\columnwidth]{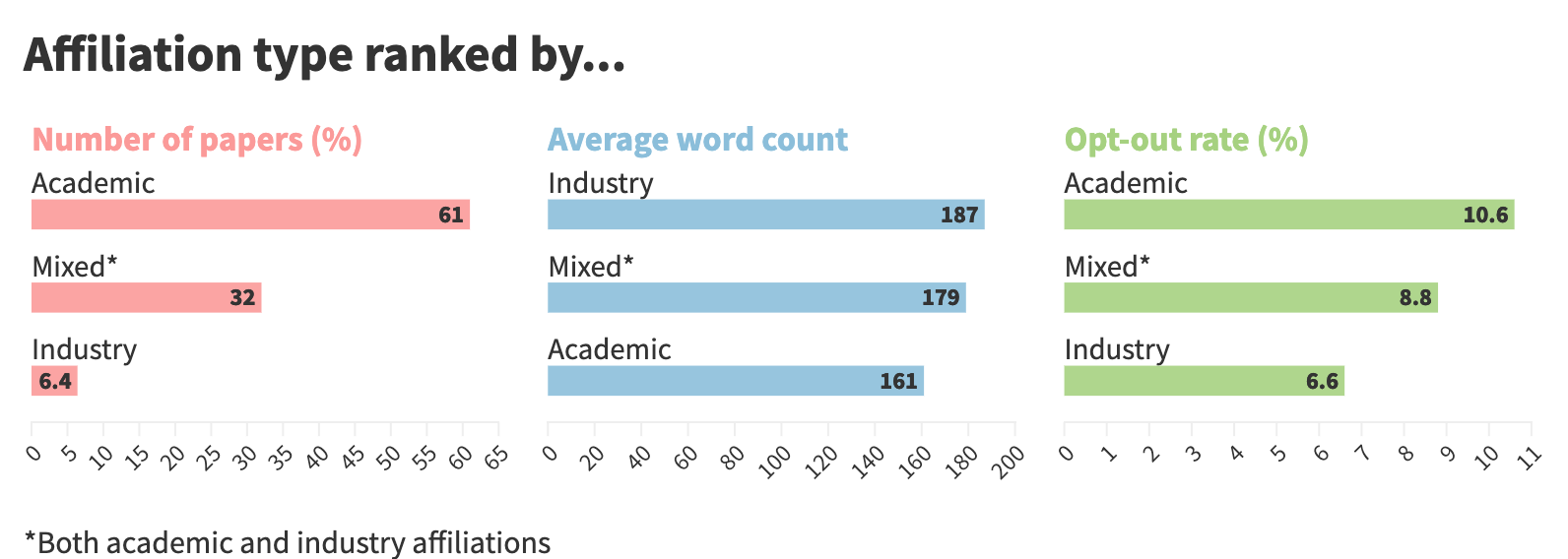}
  \caption{Affiliation type - \url{https://public.flourish.studio/visualisation/6478786}}
    \label{fig:type_sum}
\end{figure}

\begin{figure}[htbp!]
  \centering
  \includegraphics[width=0.6\columnwidth]{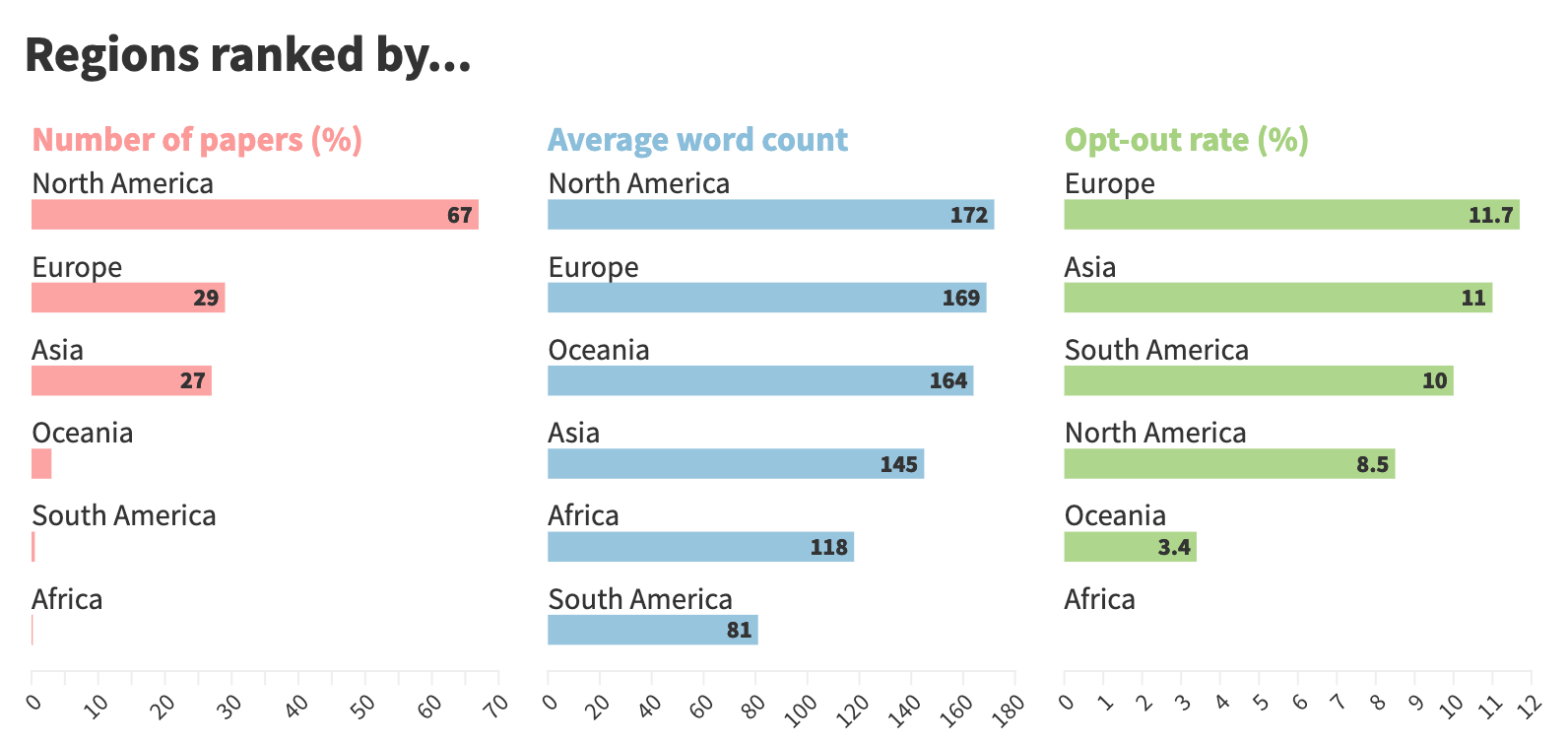}
  \caption{Regions - \url{https://public.flourish.studio/visualisation/6478761}}
  \label{fig:region_sum}
\end{figure}

We found small differences by affiliation type: opt-outs for \textit{Academic}, \textit{Industry}, and \textit{Mixed} constituted 11\%, 7\% and 9\%, though the figure for \textit{Industry} suffers from small sample size (only eight statements in the \textit{Industry} category were labelled as opt-out). See Figure ~\ref{fig:type_sum}. Results for \textit{North America}, \textit{Europe} and \textit{Asia} were similar to the average: 9\%, 12\% and 11\% respectively. See Figure ~\ref{fig:region_sum}.

\begin{figure}[htbp!]
  \centering
  \includegraphics[width=0.6\columnwidth]{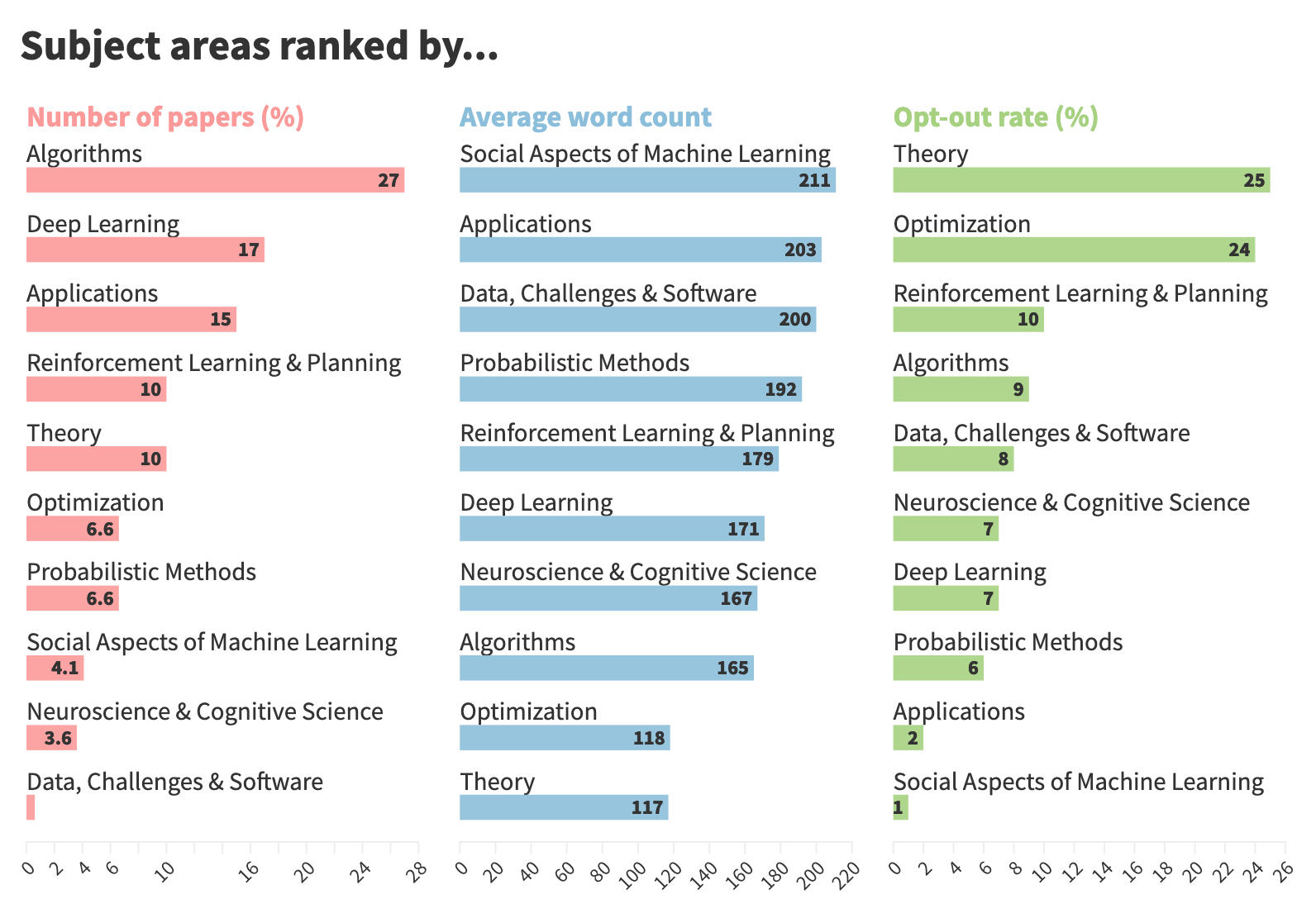}
  \caption{Primary subject area - \url{https://public.flourish.studio/visualisation/6915390}}
  \label{fig:subject_area}
\end{figure}

We found more pronounced differences in opt-out rates when disaggregated by subject area. Perhaps unsurprisingly, we found that theoretical topics, such as \textit{Theory} and \textit{Optimization} had the highest opt-out rates (25\% and 24\%), and \textit{Applications} and \textit{Social Aspects of ML} had the lowest (2\% and 1\%). See Figure ~\ref{fig:subject_area}. It is worth noting that even for \textit{Theory}, 75\% still opted in to at least some degree, despite authors of very theoretical work being told they may simply write that a statement is not applicable.

\subsubsection{Statement length}
\label{sec:length}
The mean statement length was 169 words and 7.3 sentences, i.e.\ approximately one short paragraph -- see Table ~\ref{tab-length}. This is not dissimilar to the example provided by NeurIPS in the FAQs, namely a paper with a 191 word ethics statement \citep{neurips_faq_neurips_2020}. If we look at only those papers labelled as opt-in, the mean length rises to 184 words and 7.9 sentences.

\begin{table}[htbp!]
  \caption{Average statement length}
  \label{tab-length}
  \centering
  \begin{tabular}{lll}
  \toprule
        & Mean word count     & Mean sentence count \\
    \midrule
    All papers & 169 & 7.3     \\
    All opt-in papers & 184  & 7.9      \\
    \bottomrule
  \end{tabular}
\end{table}

\begin{figure}[htbp!]
  \centering
  \includegraphics[width=0.6\columnwidth]{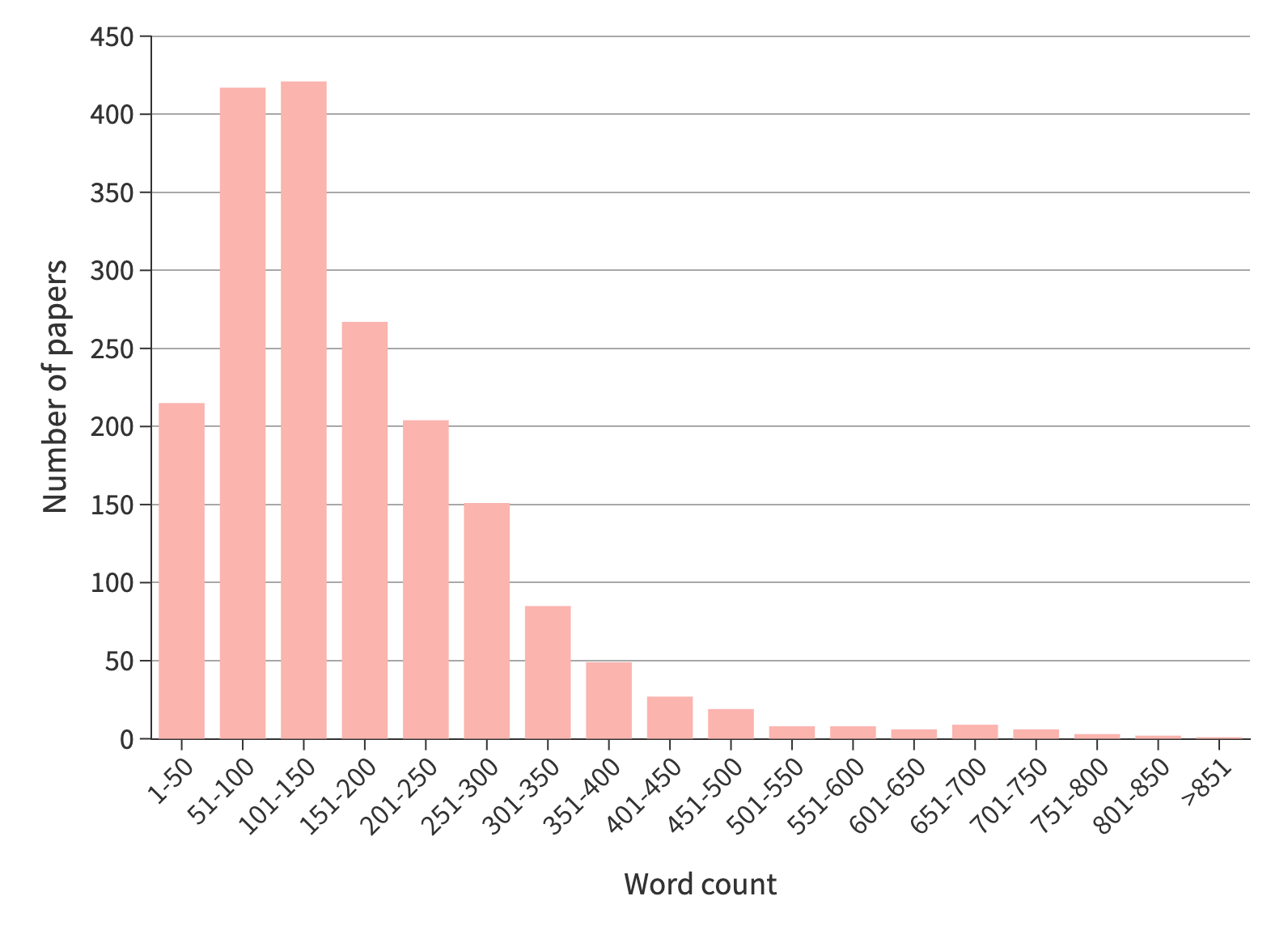}
  \caption{Length of impact statements - \url{https://public.flourish.studio/visualisation/5979712}}
  \label{fig:length}
\end{figure}

While the mean length was 169 words, there was large variation between statements. The shortest statements were two words, namely “Not applicable” (e.g.\ \citep{won2020proximity}), and the longest statement was 4337 words and 150 sentences long \citep{brown2020language}. The median length was 138 words. See Figure ~\ref{fig:length} for the entire distribution.

Again, we found small differences between affiliation type and location -- see Figures ~\ref{fig:type_sum} and ~\ref{fig:region_sum}. For example, of the 795 papers whose authors all had US affiliations, and 97 papers with Chinese affiliations, the average word lengths were 182 words and 135 words respectively.

\begin{figure}[htbp!]
  \centering
  \includegraphics[width=0.6\columnwidth]{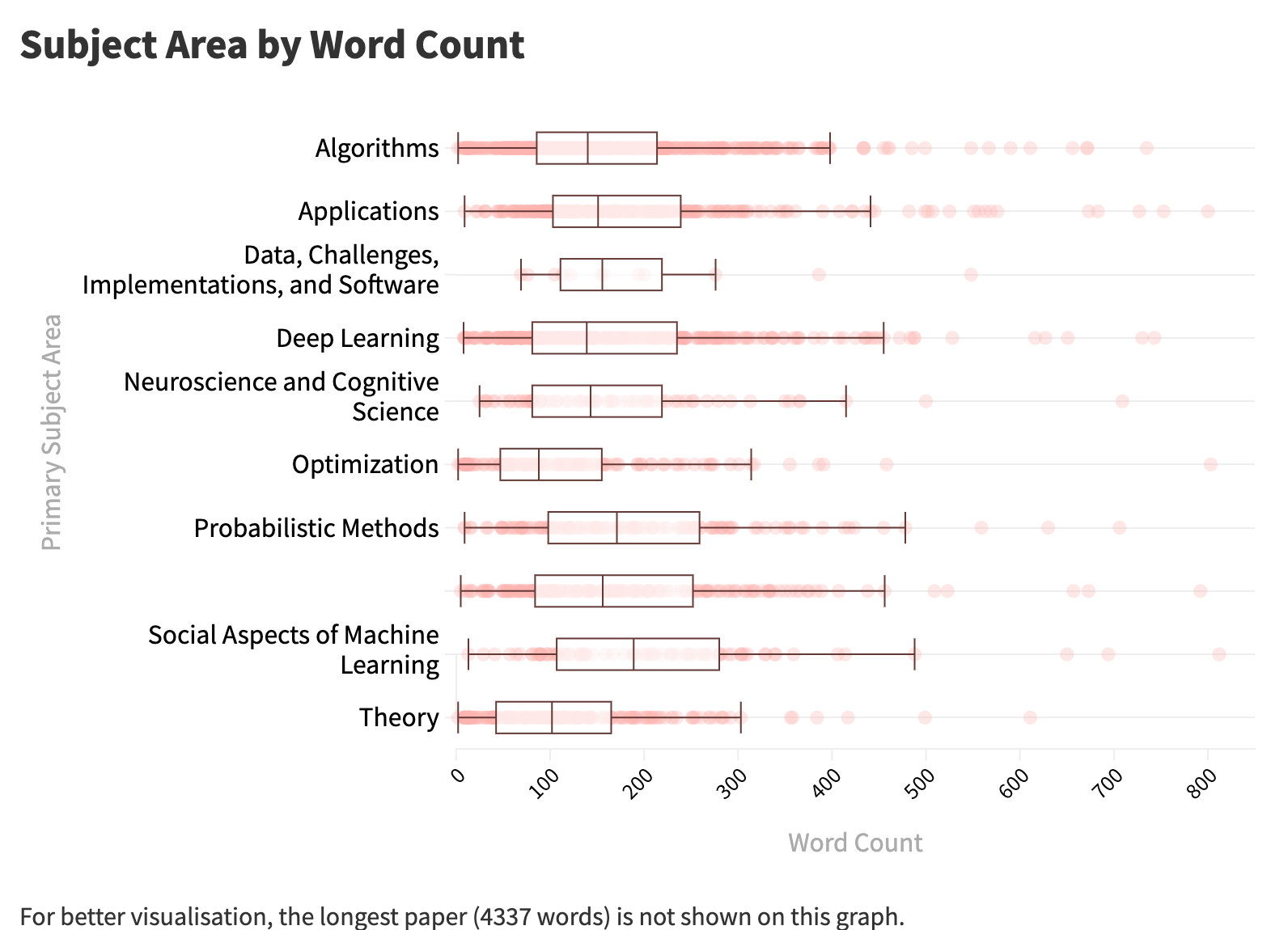}
  \caption{Length distribution by subject area - \url{https://public.flourish.studio/visualisation/6084366}}
  \label{fig:subject_area_box}
\end{figure}

Similarly to opt-out rates, we found differences between subject areas. \textit{Theory} and \textit{Optimization} had the shortest average statement length (117 and 118 words respectively), and \textit{Social Aspects} and \textit{Applications} had the longest (211 and 203 words) -- see Figures ~\ref{fig:subject_area} and ~\ref{fig:subject_area_box}.

Drilling down into the \textit{Applications} category, we can also see differences between sub-areas. In Table ~\ref{tab-applications}, we show mean word counts for all \textit{Application} sub-areas with at least 10 accepted papers. For example, we find that \textit{Health} and \textit{Computational Biology} are well above average (265 and 255 words respectively). Perhaps surprisingly, \textit{Computer Vision} and \textit{Natural Language Processing} (NLP) differ greatly (166 and 340 words respectively). However, we note that the longest statement (4337 words) is found in an NLP paper, namely OpenAI’s GPT3 paper \textit{Language models are few-shot learners} \citep{brown2020language}. The average word count for NLP statements excluding this paper is 223 words -- around a third longer than those from \textit{Computer Vision}.

\begin{table}[htbp!]
  \caption{Application sub-areas}
  \label{tab-applications}
  \centering
  \begin{tabular}{lll}
  \toprule
    Primary subject area    & \multicolumn{1}{p{1cm}}{\raggedright Num \\ papers }     & \multicolumn{1}{p{1cm}}{\raggedright Mean\\ word \\ count } \\
    \midrule
    Natural Language Processing & 35 & 340  \\
    Health & 12 & 265 \\
    Computational Biology and Bioinformatics & 12 & 255 \\
    Network Analysis & 11 & 236 \\
    Time Series Analysis & 10 & 219 \\
    Applications [No sub-topic listed] & 15 & 170 \\
    Computer Vision & 96 & 166 \\
    \bottomrule
  \end{tabular}
\end{table}

\subsection{Themes}
\label{sec:themes}
For a thorough treatment of the themes covered in a large sample of impact statements, see \citet{nanayakkara2021unpacking}. To get a flavour of some of the topics discussed in all statements, and whether these differed by affiliation type and subject area, we enumerated the highest frequency words across all impact statements, with stopwords (such as ``the'' and ``and'') removed.

\begin{table}[htbp!]
  \captionof{table}{Most common words}
  \label{tab:top}
  \centering
  \begin{tabular}{ll|ll}
  \toprule
    Word     & Freq  & Word & Freq \\
    \midrule
    learning &    2516   &  method &    816      \\
    work &    2257   &  methods &    789   \\
    data &    1941   &  paper &    786   \\
    models &    1393   &  neural &    649   \\
    applications &    1387   &  many &    647   \\
    model &    1077   &  deep &    635   \\
    research &    1022   &  new &    611   \\
    machine &    945   &  networks &    588   \\
    systems &    886   &  theoretical &    588   \\
    used &    876   &  use &    565   \\
    will &    875       &  proposed &    555   \\
    impact &    872   &  one &    553   \\
    algorithms &    845   &  eg &    524   \\
    potential &    840   &  tasks &    510   \\
    training &    827  &  societal & 497\\
    \bottomrule
  \end{tabular}
\end{table}

The top 30 words can be found in Table ~\ref{tab:top}. As illustrated by the top five words (\textit{learning}, \textit{work}, \textit{data}, \textit{models}, \textit{applications}), the list contains a high proportion of ``technical'' terms. In Table ~\ref{tab-top-societal}, we therefore give a list of the top 30 words that we deemed to be related to societal impacts. We show in bold those terms related to particular classes of societal considerations.

\begin{table}[htbp!]
  \captionof{table}{Top words relating to societal impacts}
  \label{tab-top-societal}
  \centering
  \begin{tabular}{ll|ll}
  \toprule
    Word     & Freq & Word & Freq   \\
    \midrule
    impact &    872     &  community & 272   \\
    societal &  497     &  researchers &     271   \\
    human &    415     &  \textbf{robustness} &    260   \\
    \textbf{privacy} &    403     &  broader &    257   \\
    future &    398     &  risk &    254   \\
    \textbf{biases} &    334     &  \textbf{fairness} &    243   \\
    realworld &     317     &  study &    241   \\
    consequences &     315     &  autonomous &    240   \\
    \textbf{robust} &     309     &  medical &    238   \\
    \textbf{bias} &     307     &  real &    236   \\
    social &     296     &  users &    224   \\
    \textbf{adversarial} &     296     &  world &    219   \\
    society &     291     &  present &    213   \\
    ethical &     289     &  people &    206   \\
    impacts &     282     &  \textbf{safety} &    202   \\
    \bottomrule
  \end{tabular}
\end{table}

This list gives an indication of some of the most common societal considerations considered, namely privacy (\textit{privacy} occurred 403 times), fairness (\textit{fairness}, \textit{biases}, \textit{bias} occurred 243, 334, 307 times respectively), robustness (\textit{robust}, \textit{robustness}, \textit{adversarial} occurred 309, 260, 296 times), and safety (\textit{safety} occurred 202 times). These themes were fairly consistent across locations and affiliation types. We found the top words associated with application settings were \textit{medical} (238 occurrences), \textit{robots} (123), \textit{science} (123), \textit{malicious} (113), \textit{scientific} (106), \textit{healthcare} (104), \textit{health} (96), \textit{robots} (94), \textit{decision-making} (79), \textit{surveillance} (79) and \textit{industry} (73).

\subsection{Valence}
\label{sec:val}
To get an indication of whether authors described both positive and negative impacts to the same extent, we created a list of synonyms for \textit{positive} using Wordnet's synset (semantic) relations \citep{wordnet}. We took the number of occurrences of positive synonyms that were not negated as a measure of explicitly positive sentiment. We repeated this for synonyms of \textit{negative}, \textit{strength} and \textit{limitation} -- see Table ~\ref{tab-valence}. This method has many limitations, including that it does not account for implicit positive impacts. For example, if an author states their work could be applied to healthcare, they may intend for this to be viewed as a positive impact, though no synonym for positive is used. The counts should therefore be taken to be a crude proxy.

From Table ~\ref{tab-valence} we see that the number of occurrences of synonyms for \textit{positive} is higher than \textit{negative} (an average of 4.6 occurrences per statement versus 3.6). This quantitative measure supports the observations about valence given by \citet{nanayakkara2021unpacking}, who describe that some researchers omit discussions of negative impacts, explicitly state that there are no negative impacts, or only include a brief mention of negative impacts. We note a similar bias towards occurrences of \textit{strength} synonyms, over \textit{limitation} synonyms (1.3 versus 0.6 per statement).

\begin{table}[htbp!]
  \caption{Valence}
  \label{tab-valence}
  \centering
  \begin{tabular}{lll}
  \toprule
    Synonyms    & Mean occurrences     & \% statements \\
    \midrule
    Positive & 4.6 & 90\%     \\
    Negative & 3.6 & 80\%     \\
    \midrule
    Strength & 1.3 & 60\%     \\
    Limitation & 0.6 & 37\%     \\
    \bottomrule
  \end{tabular}
\end{table}

\section{Discussion}
\label{sec:discussion}

In our analysis, we have found that there is a concentration of authors along geographic and institutional lines. We find high variation in engagement, as measured by length and opt-out rates. We find evidence that certain established topics are common to impact statements, namely privacy, fairness, robustness and safety. We also find evidence that authors tend to discuss positive impacts to a greater extent than negative impacts. 
We now reflect on what these tell us about the benefits and challenges of broader impact requirements. We use the framing from \citet{prunkl2021institutionalizing}, in which potential benefits, negative outcomes and challenges were identified based on lessons learnt from related governance mechanisms. 

\subsection{Evidence of benefits}
\label{sec:benefits}

Building on the EPSRC AREA framework, \citet{prunkl2021institutionalizing} categorise the potential benefits of broader impact statements under the headings of \textit{Anticipation}, \textit{Action}, \textit{Reflection and awareness}, and \textit{Coordination}. We find evidence relating to the latter two categories as follows.

\paragraph{Reflection and awareness} Most authors (90\%), including the majority of those whose primary subject area falls under theoretical areas such as \textit{Theory} (75\%), took the opportunity to reflect on their work and include an impact statement, rather than stating that one was not applicable (\S~\ref{sec:opt-outs}). In a survey of NeurIPS authors, \citet{abuhamad2020like} found that of the respondents who supported the requirement, some found the thought process to be the most valuable aspect, suggesting that the opportunity for reflection is valued by some. The long tail of statement lengths shows that several authors took the unlimited page limit as an opportunity to include a thorough treatment of some of the issues they had identified, raising more detailed awareness of such issues among their readers (\S~\ref{sec:length}). For example, in \textit{Language models are few-shot learners} (a winner of the NeurIPS 2020 Best Paper Awards), the impact section includes detailed discussions of potential misuse, fairness, and energy use \citep{brown2020language}. This paper contained the longest impact statement, and included authors whose main contribution to the paper was detailed analysis of these ethical considerations. For an additional discussion of encouraging trends found in preprint versions of NeurIPS papers, see \citet{boyarskaya2020overcoming}.

\paragraph{Coordination} These statements have given an indication of some of the issues which many authors felt applied to their work, such as privacy, fairness, robustness and safety (\S~\ref{sec:themes}). This correlates with the themes found by \citet{nanayakkara2021unpacking}, who identified the following issues as themes amongst their sample: bias (24\%), robustness and reliability (21\%), privacy (19\%), environment (10\%), interpretability (10\%) and labor (6\%). Identifying these common themes provides some information about which issues are widely recognised (or otherwise) within the ML community.

Another potential coordinating function of statements is prompting collaboration with other disciplines or stakeholders. While this is difficult to ascertain from the statements directly, \citet{abuhamad2020like} found that 16\% of survey respondents reached out for outside support. Indeed \citet{brown2020language} used a contribution section to highlight individuals who worked on fairness analysis, threat analysis and ethical impacts (including the broader impact section).

\subsection{Evidence of negative outcomes}
\label{sec:negatives}

\citet{prunkl2021institutionalizing} list quality deficits, trivialisation of ethics and governance, negative attitudes, false sense of security, unintended signalling, and polarisation of the research community as potential negative outcomes of impact statement requirements. Our analysis finds evidence for the first two of these.

\paragraph{Quality deficits} As previously discussed, we found high variation in engagement, as measured by length (\S~\ref{sec:length}) and opt-out rates (\S~\ref{sec:opt-outs}), as well as a bias towards discussing positive aspects (\S~\ref{sec:val}). The average statement is relatively short, around seven sentences (\S~\ref{sec:length}). \citet{abuhamad2020like} found that the vast majority of authors surveyed spent less than two hours on their statement, and just under half spent less than one hour. Some NeurIPS referees remarked on the low quality of statements on social media \citep{prunkl2021institutionalizing}. Some authors did not engage with negative impacts at all, and some only focused on impacts to their technical field, rather than societal impacts \citep{nanayakkara2021unpacking}. 

\paragraph{Trivialisation of ethics} Relatedly, one might interpret the large number of very short statements as evidence of the trivialisation of the societal impacts of NeurIPS work.\\

Other negative outcomes, such as negative attitudes and polarisation are difficult to gauge directly from the statements themselves; other means such as attitude surveys and deliberative fora are better suited. There was certainly some degree of backlash expressed on social media, and it is possible that negative attitudes towards the requirement was a contributing factor to changing the process for 2021 (the chairs refer to author feedback as one factor leading to the decision \citep{intro_checklist}).

\subsection{Causes and challenges}
\label{sec:challenges}
Here we discuss several challenges relating to impact statements, and potential causes of the negative outcomes identified. Possible causes for quality deficits include lack of explanation and guidance; the complexity of the task (particularly for foundational and general purpose research); high opportunity costs; and institutional, social, cognitive pressure and biases \citep{prunkl2021institutionalizing}.

\paragraph{Institutional, social and cognitive bias} The original motivation given by the NeurIPS chairs was that it was incumbent on the community ``to consider not only the beneficial applications and products enabled by our research, but also potential nefarious uses and the consequences of failure'' in response to the ``more and more pervasive'' impact of the community's work \citep{NeurIPS_getting_2020}. While it was stated that authors must ``take care to discuss both positive and negative outcomes'', we find evidence that authors tend to discuss positive impacts to a greater extent, with some authors neglecting negative impacts altogether (\S~\ref{sec:val}). One possible contributing factor is that researchers may be incentivized to focus on minor risks or those that do not threaten their own or organisational interests \citep{prunkl2021institutionalizing}.

Because of the high concentration of papers associated with a handful of affiliations (\S~\ref{sec:top_aff}), any institutional bias from the most prolific institutions is likely to have a significant effect. Many academic researchers are likely to be affected by industry incentives, not least because of the large number of papers that are collaborations between authors with academic and industry affiliations (\S~\ref{sec:aff_type}); see also \citep{hagendorff2021ethical}.

The concentration of authors along geographic lines (\S~\ref{sec:loc}) is also likely to influence how authors view the impacts of their work. We observed different levels of engagement along geographic lines, for example between the US and China (\S~\ref{sec:loc}), which could be a result of differing attitudes towards ethics and the societal impact of ML \citep{roberts2021chinese}. To date, AI ethics in China has been more government driven than company driven (for example, the government recently released new high-level AI ethics guidelines \citep{shen_chinese_2021}), whereas American Big Tech companies tend to follow their own codes of ethics in absence of specific government guidelines. This could mean that longer statements from US companies are a result of their companies efforts in this space (which could be in good faith, or could suffer from ethics washing). Language could of course be another contributing factor. Since NeurIPS papers must be written in English we might expect typically longer discussions from those for whom English presents a lower language barrier, which may be more prevalent among US affiliated authors than those affiliated with institutions in China.

\paragraph{Lack of best practice, guidance and explanation of purpose} Although some more detailed unofficial guidance was made available, such as \citet{ashurst_guide_2020} and \citet{hecht_suggestions_2020}, the \textit{official} guidance given was very limited. A potential cause of the predominance of positive over negative impacts discussed is that authors confused the requirement with other broader impact requirements. For example, the wording surrounding broader impact statements that form part of funding applications often focus on highlighting positive impacts \citep{prunkl2021institutionalizing}. It is clear that different authors interpreted the aim of the requirement very differently. For example, some only discussed positive societal impacts, some restricted to technical impacts to their field, some focused solely on whether there were new societal impacts introduced by their work \citep{moulos2020finite}, some on whether there were immediate impacts (e.g.\ within the next 6 months \citep{wu2019adaptive}) and others focused on other aspects of responsible research such as reproducibility \citep{colas2020language}.

Conversely, some authors stuck rigidly to the (limited) guidance given. For example, some closely followed the four bullets suggested in the template file \citep{neurips_template_2020, wu2020graph}. This shows the importance of careful guidance.

It is also possible that lack of communication around the purpose of the requirement (and why it might be valuable) contributed to push back from some researchers. Introducing a new requirement that has associated costs (even a fairly lightweight and flexible requirement, as was arguably the case here) will only be received well if its benefits are understood.

\paragraph{Opportunity costs} 
Impact statements did not contribute to the page limit, thus removing one potential opportunity cost. This allowed authors to include extended and thorough statements if they so desired. We note that this is no longer the case for 2021 papers: any discussion of negative societal impact counts towards the page limit. While an additional page has been given to account for this (and other new requirements), there is nothing to prohibit authors from using the additional page for other uses, and so any discussion of impacts is in competition with these other uses. There is, however, a downside to the 2020 approach of having a separate section after the page limit: having an impact statement after the main body does separate it from the central work. Some would argue that an integrated approach is more desirable, since societal reflection should be an activity integrated with the research itself.

Additionally, writing a high-quality statement can take time. The high number of very short impact statements, and the short amount of time spent on statements \citep{abuhamad2020like}, may be at least partly a result of this opportunity cost.

\subsection{Conclusion}

\subsubsection{Summary}
The 2020 NeurIPS impact statements present a unique opportunity to investigate the benefits and challenges of this and similar governance mechanisms, and also provide an insight into how ML researchers think about the societal impacts of their work. In order to encourage investigations into how researchers responded to this requirement, we have created a dataset containing the impact statements from all NeurIPS 2020 papers, along with additional information such as affiliation type, location and subject area. We also provide a visualisation tool for exploration and an initial quantitative analysis of the dataset. We investigate the voices represented (\S~\ref{sec:voices}), levels of engagement (\S~\ref{sec:engagement}), themes (\S~\ref{sec:themes}) and valence (\S~\ref{sec:val}), and discuss how these reflect the benefits (\S~\ref{sec:benefits}), negative outcomes (\S~\ref{sec:negatives}) and challenges (\S~\ref{sec:challenges}) associated with broader impact statements. We encourage others to investigate the dataset of impact statements, and to continue to further analyse how researchers respond to changing requirements, such as the checklist approach, over the coming years. 

\subsubsection{What we can learn from 2020 statements}

The 2020 requirement provides lessons for self-governance mechanisms more broadly, in particular:

\paragraph{The importance of creating the right incentives} While the requirement was mandatory, there were few incentives to engage deeply with the task, nor any requirement that statements meet any particular standard of quality. While some opportunity costs were accounted for (namely page length), in many ways the process incentivised minimal engagement. For any self-governance mechanism, one must carefully consider how individuals are incentivised to act, and whether this is likely to result in the desired aims.

\paragraph{The importance of clear expectations and guidance} As discussed, there may have been confusion between this requirement and other impact statement requirements which focus on potential benefits. The range of approaches taken by authors show that the aim and expectations were not universally understood. Even within the limited guidance given, it was not clear which elements of responsible research were being targeted. There is therefore a need to disentangle the different elements of responsible research (such as reproducibility, protections for human and data subjects, and anticipation of future downstream consequences), to target mechanisms specifically, and to be clear about the aims. On this occasion, outside guidance did provide some additional help for researchers (\citep{ashurst_guide_2020},\citep{hecht_suggestions_2020}); clearer communication of the aims and expectations also better enable external researchers to provide appropriate supporting materials such as these.

\paragraph{The importance of transparency, and constructive deliberation} The wide range of approaches, levels of engagement, and responses to the requirement show a continued lack of consensus around the purpose of mechanisms such as impact statements. Continued deliberation within the community, in consultation with impacted stakeholders and societal experts therefore continues to be of great importance.\\

More specifically, what can the 2020 statements tell us about the utility of impact statement requirements? Unfortunately, it is challenging to judge the full potential of impact statement requirements based on one year alone, particularly when little guidance (and warning) was available to researchers. It takes time to build up best practice, good guidance, and effective incentives. We have identified a range of challenges which could be addressed over time to improve outcomes if such a requirement were to persist. However, we have already seen some evidence of the benefits from just one year of the requirement: including providing the opportunity for some researchers to engage thoughtfully, and the opportunity to include extended statements that thoroughly discuss a range of concerns to be addressed. We found that most authors (including those in the most theoretical categories) showed some degree of engagement, promoting reflection and raising awareness of some of the potential risks. Regarding potential negative outcomes, many of these need to be measured through other means, such as attitude surveys. We did find evidence that many statements had limited engagement with the requirement, with some failing to acknowledge any negative consequences. Again, to judge the full potential, this needs to be measured over time. Although monitoring how researchers address societal impacts in their papers may be more challenging under the new checklist requirement (since societal impacts are not required to be in a separate clearly labelled section, but may be discussed throughout the paper), we urge the community to continue to track how researchers choose to do this.\\

Our hope is that as the ML community continues to grapple with its responsibilities and experiment with different governance mechanisms to encourage responsible research, we will move towards becoming a more mature field with respect to ethical ML. We would like to see societal thinking become a more integrated part of ML research, informing which research is undertaken, and how it is executed. This is of particular importance for research towards the application end of the spectrum, and for research involving data generated by or about people. Until then, we hope the community will (i) continue to experiment with and test governance mechanisms and their associated incentives, (ii) continue to engage in open deliberation both within the community and with outside engagement, and (iii) continue to raise awareness of identified potential harms. While it is encouraging to see that certain classes of harms have become relatively widely recognised (such as privacy and bias), we hope the community will increase its awareness and understanding of a wider range of harms and issues.

\begin{acks}
For discussion and input we thank: Markus Anderljung, Jeff Ding, Ben Garfinkel, Matthij Maas, Carina Prunkl and Toby Shevlane.
We also thank Earl Ng for the code to extract the impact statements \citep{pdf-scraper}, and for providing code review.
\end{acks}

\bibliographystyle{ACM-Reference-Format}
\bibliography{biblio}

\end{document}